\documentclass[aps,twocolumn,showpacs,floatfix]{revtex4}
\usepackage{graphicx}
\usepackage{amsmath,amsfonts}
\usepackage{amsbsy}

\begin{document}

\title{Equilibration of a one-dimensional quantum  liquid}

\author{K. A. Matveev}

\affiliation{Materials Science Division, Argonne National Laboratory,
  Argonne, Illinois 60439, USA}

\date{April 8, 2013}

\begin{abstract}

  We review some of the recent results on equilibration of
  one-dimensional quantum liquids.  The low-energy properties of these
  systems are described by the Luttinger liquid theory, in which the
  excitations are bosonic quasiparticles.  At low temperatures the
  relaxation of the gas of excitations toward full equilibrium is
  exponentially slow.  In electronic Luttinger liquids these
  relaxation processes involve backscattering of electrons and give
  rise to interesting corrections to the transport properties of
  one-dimensional conductors.  We focus on the phenomenological theory
  of the equilibration of a quantum liquid and obtain an expression
  for the relaxation rate in terms of the excitation spectrum.

\end{abstract}


\maketitle

\section{Introduction}
\label{sec:introduction} 

The low energy properties of one-dimensional quantum systems are
commonly described in the framework of the so-called
Tomonaga-Luttinger liquid \cite{tomonaga1950remarks,
  luttinger_exactly_1963, mattis_exact_1965,
  dzyaloshinskii1974correlation, luther_single-particle_1974,
  mattis_new_1974, efetov_correlation_1975, haldane1981luttinger,
  giamarchi2004quantum}.  This description applies to systems of
either bosons or fermions, but regardless of the statistics of the
constituent particles, the excitations of the Luttinger liquid are
bosons with linear spectrum.  The Hamiltonian of a Luttinger liquid is
given by
\begin{equation}
  \label{eq:Luttinger_Hamiltonian}
  H=\sum_q  v|q|b_q^\dagger b_q^{} + 
           \frac{\pi\hbar}{2L}[v_N (N-N_0)^2+v_J J^2],
\end{equation}
where $b_q$ is the annihilation operator of a bosonic excitation with
momentum $q$ propagating with velocity $v$,
\cite{haldane1981luttinger}.  The Hamiltonian
(\ref{eq:Luttinger_Hamiltonian}) assumes that the system has a finite
size $L$ and periodic boundary conditions are imposed.  Apart from
the occupation numbers of bosonic states, the energy of the system
depends on two integer numbers, $N$ and $J$.  In the case of fermionic
Luttinger liquids they can be interpreted in terms of the numbers of
right- and left-moving particles as $N=N^R+N^L$ and
$J=N^R-N^L$.  Parameters $v_N$ and $v_J$ have dimension of velocity
and depend on the interactions between the particles; $N_0$ is some
reference number of particles in the system.
The Luttinger liquid theory described by the Hamiltonian
(\ref{eq:Luttinger_Hamiltonian}) has been successful in predicting a
number of interesting phenomena, such as the renormalization of
impurity scattering in interacting one-dimensional electron systems
\cite{kane_transmission_1992, furusaki_single-barrier_1993}
subsequently observed in experiments \cite{tarucha_reduction_1995,
  auslaender_experimental_2000, bockrath_luttinger-liquid_1999,
  yao_carbon_1999}.  

An interesting feature of the model (\ref{eq:Luttinger_Hamiltonian})
is the complete absence of coupling between the bosons.  As a result,
the lifetimes of bosonic excitations are infinite and the system does
not relax toward thermal equilibrium.  It is important to keep in
mind, however, that Eq.~(\ref{eq:Luttinger_Hamiltonian}) is the exact
Hamiltonian of the system only for the so-called Luttinger model
\cite{luttinger_exactly_1963} where the spectrum of the fermions
consists of two linear branches, $\epsilon_p=\pm v_F p$.  In a generic
situation this is an approximation applicable only in the vicinity of
the two Fermi points, and thus the Luttinger theory
(\ref{eq:Luttinger_Hamiltonian}) applies only at low energies.  In
other words, Eq.~(\ref{eq:Luttinger_Hamiltonian}) represents a fixed
point Hamiltonian in the renormalization group sense, which should, in
principle, be amended by additional contributions describing various
irrelevant perturbations.  The latter are the operators of third and
higher powers in $b_q$ and $b_q^\dagger$, which result in scattering
of bosonic excitations.  They adequately account for the curvature of
the spectrum near the Fermi points, which gives rise to a multitude of
interesting phenomena studied in the last few years.  (See
Ref.~\onlinecite{imambekov_one-dimensional_2012} for a recent review.)

Another important aspect of the Luttinger model is that the original
fermions are classified as belonging to one of two species: the right-
and left-moving particles.  In realistic systems there is no
fundamental difference between the particles moving in opposite
directions, and a right-moving fermion may become a left-moving one
upon scattering.  These backscattering processes give rise to several
interesting phenomena not captured by the Luttinger liquid theory.

One example is the effect of backscattering on the
transport properties of the system.  Experimentally transport can be
studied in quantum wire devices \cite{van_wees_quantized_1988,
  wharam_one-dimensional_1988}, where a one-dimensional system is
smoothly connected to two-dimensional leads.  In the absence of
interactions, the conductance of a quantum wire is quantized in units
of $e^2/h$, where $e$ is the elementary charge and $h$ is the Planck's
constant.  Interactions between electrons included into the Luttinger
liquid theory do not affect conductance quantization
\cite{maslov_landauer_1995, ponomarenko_renormalization_1995,
  safi_transport_1995}.  On the other hand, the backscattering
processes excluded from model (\ref{eq:Luttinger_Hamiltonian}) reduce
the conductance \cite{lunde_three-particle_2007}.  More detailed
theories of conductance of long uniform quantum wires relate the
correction to conductance due to electron-electron interactions to the
rate of equilibration of the electron liquid
\cite{micklitz_transport_2010, matveev_equilibration_2011}.

The physics of equilibration of a liquid of one-dimensional fermions
is the main subject of this paper.  It is another example of a problem
where backscattering processes are crucial.  For particles with a
realistic spectrum, such as $\epsilon_p=p^2/2m$, the relaxation of the
system to equilibrium involves backscattering processes changing the
numbers $N^R$ and $N^L$.  On the other hand, in a Luttinger liquid the
difference $J=N^R-N^L$ is conserved, even if the irrelevant
perturbations are taken into account.  As a result, equilibrium states
of the system described by the Hamiltonian
(\ref{eq:Luttinger_Hamiltonian}) are characterized by different
chemical potentials of the two species of particles, $\mu^R$ and
$\mu^L$, and their relaxation to a single value $\mu$ is neglected.

Below we discuss the mechanism of equilibration of one-dimensional
quantum liquids beyond the Luttinger liquid approximation.  An
expression for the corresponding equilibration rate $\tau^{-1}$ was
obtained microscopically for the regimes of both weak
\cite{micklitz_transport_2010} and strong
\cite{matveev_equilibration_2010} interactions.  An alternative
phenomenological approach \cite{matveev_equilibration_2012,
  matveev_scattering_2012} based on the Luttinger liquid theory is
applicable at any interaction strength and results in an expression
for the equilibration rate $\tau^{-1}$ in terms of the excitation
spectrum of the system.  The latter can be either measured
experimentally or derived microscopically for specific models.  In
Secs.~\ref{sec:equil-state}--\ref{sec:discussion} we review the
phenomenological approach \cite{matveev_equilibration_2012,
  matveev_scattering_2012} and discuss the implications of the results
for the equilibration rate to experiments with quantum wires.

\section{Equilibrium state of a uniform Luttinger liquid}
\label{sec:equil-state}

Let us first discuss the possible equilibrium states of a Luttinger
liquid.  In general, the equilibrium distribution is determined by the
integrals of motion of the system \cite{landau_statistical_1980}.  We
will assume that the irrelevant perturbations resulting in weak
scattering of bosons are added to the Hamiltonian
(\ref{eq:Luttinger_Hamiltonian}).  Then there are four integrals of
motion: energy, momentum, and the numbers of right- and left-moving
particles, $N^R=(N+J)/2$ and $N^L=(N-J)/2$.  The Gibbs probability of
realization of a given many-particle state $i$ is then given by
\begin{equation}
  \label{eq:Gibbs}
  w_i=\frac{1}{Z}\exp
      \left(
        -\frac{E_i+uP_i-\mu^LN^L-\mu^RN^R}{T}
      \right),
\end{equation}
where $E_i$ and $P_i$ are the values of the momentum of the system in
state $i$.  To obtain the equilibrium distribution of Bose excitations
one also needs the expression \cite{haldane1981luttinger} for the
momentum of the Luttinger liquid
\begin{equation}
  \label{eq:Luttinger_Momentum}
  P=p_F J +\sum_q  q\, b_q^\dagger b_q^{},
\end{equation}
where the Fermi momentum is defined via particle density,
$p_F=\pi\hbar N/L$.

Using the expression (\ref{eq:Gibbs}) one easily obtains the
equilibrium form of the occupation numbers of the boson states:
\begin{equation}
  \label{eq:Boson_distribution}
  N_q=\frac{1}{e^{(v|q|-uq)/T}-1}.
\end{equation}
Note that as a result of momentum conservation the Bose distribution
depends not only on temperature but also the parameter $u$, which can
be thought of as the velocity of the gas of bosonic excitations.  

In addition to the bosonic occupation numbers, the state of the liquid
depends on the zero modes $N$ and $J$.  According to
Eq.~(\ref{eq:Gibbs}) in thermal equilibrium the latter is peaked
sharply near an average value $J$, which satisfies
\begin{equation}
  \label{eq:J_condition}
  \pi\hbar\frac{v_J J}{L}=up_F+\frac{1}{2}\Delta\mu,
\end{equation}
where $\Delta\mu=\mu^R-\mu^L$.  In a Luttinger liquid the ratio
$j=v_JJ/L$ has the meaning of the particle current
\cite{haldane1981luttinger}.  Expressing the latter in terms of the
drift velocity $v_d$ as $j=(N/L)v_d$, we find
\begin{equation}
  \label{eq:u-v_d_relation}
  v_d=u+\frac{\Delta\mu}{2p_F}.
\end{equation}
This expression shows that in an equilibrium state of the Luttinger
liquid the gas of excitations moves at a velocity $u$ different from
the velocity $v_d$ of the system as a whole.  This decoupling is a
result of conservation of $J$, which allows for the possibility of
$\Delta\mu\neq0$.  In a realistic system the backscattering processes
bring about relaxation of $\Delta \mu$ to zero, and the velocities $u$
and $v_d$ equilibrate.

\section{Equilibration rate}
\label{sec:equilibration-rate}

In order to study the kinetics of equilibration of a Luttinger liquid
one has to consider the corrections to the fixed point Hamiltonian
(\ref{eq:Luttinger_Hamiltonian}).  In the case of spinless Luttinger
liquid the irrelevant perturbations are terms of third and higher
orders in bosonic operators, such as $b_{q_1+q_2}^\dagger
b_{q_1}b_{q_2}$, $b_{q_1+q_2-q_3}^\dagger b_{q_3}^\dagger b_{q_1}
b_{q_2}$, etc.  Such perturbations give rise to scattering of the
bosonic excitations and to relaxation of their distribution function
toward the equilibrium distribution (\ref{eq:Boson_distribution}).
Since the scattering of bosons conserves their total momentum, the
resulting distribution is characterized by a velocity $u$, which can
be easily obtained from the initial momentum of the whole gas of
excitations.  When the distribution approaches the equilibrium form
(\ref{eq:Boson_distribution}) the typical scattering events involve
bosons with energies of order temperature, and the scattering rate
$\tau_0^{-1}$ scales as a power of $T$.  For instance, in the case of
a strongly interacting system the equilibration rate of the gas of
excitations scales at $\tau_0^{-1}\propto T^5$
\cite{lin_thermalization_2013}.

\begin{figure}
\includegraphics[width=.3\textwidth]{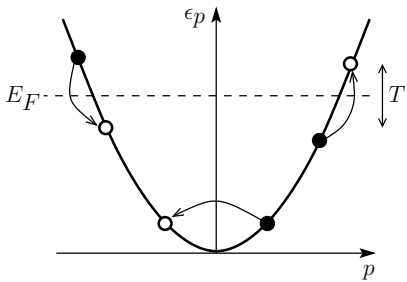}
\caption{In the model of weakly interacting fermions the dominant
  backscattering process involves three particles: one near the bottom
  of the band and the other two near the left and right Fermi points
  \cite{lunde_three-particle_2007}.}
\label{fig:lunde}
\end{figure}

The backscattering processes required for the relaxation of the
velocity $u$ toward $v_d$ have been studied microscopically in the
regime of weak interaction in
Ref.~\onlinecite{lunde_three-particle_2007}.  Because of the
constraints imposed by the conservation of momentum and energy the
simplest non-trivial process involves three particles,
Fig.~\ref{fig:lunde}.  The Fermi statistics requires that in the
dominant backscattering process two particles are within the energy
range of order temperature from the left and right Fermi points,
whereas the third one is within $T$ from the bottom of the band.  As a
result of such a scattering event the third particle backscatters,
i.e., the numbers of right- and left-moving particles change by one.
Since the backscattering particle fills a hole deep below the Fermi
level, the rate of such processes is exponentially small,
$\tau^{-1}\propto e^{-E_F/T}$ \cite{lunde_three-particle_2007,
  micklitz_transport_2010}.  We will see below that the backscattering
rate is exponentially suppressed at low temperatures for any
interaction strength.

The strong suppression of the backscattering rate means that at low
temperatures the equilibration of the quantum liquid proceeds in two
steps.  First, the bosonic excitations come to thermal equilibrium
with each other and their distribution function takes the form
(\ref{eq:Boson_distribution}).  This thermalization takes a relatively
short time of order $\tau_0$.  Second, over much longer time $\tau$
the backscattering processes equilibrate the zero mode $J$ with the
bosons.  During this time the velocity $u$ of the gas of bosonic
excitations approaches the velocity $v_d$ of the liquid as the
difference of the chemical potentials of the right and left movers
$\Delta \mu$ relaxes to zero, see Eq.~(\ref{eq:u-v_d_relation}).  The
time dependences of $u$ and $\Delta\mu$ should follow the usual
relaxation law
\begin{equation}
  \label{eq:relaxation}
  \frac{du}{dt} = -\frac{u-v_d}{\tau}, 
\quad 
  \frac{d}{dt}\Delta\mu=-\frac{\Delta\mu}{\tau}.
\end{equation}
Expression (\ref{eq:relaxation}) gives the formal definition of the
equilibration time $\tau$.

In order to study the relaxation rate $\tau^{-1}$ at arbitrary
interaction strength it is tempting to use the Luttinger liquid
description of the system.  However, this approach is incapable of
describing the particles near the bottom of the band,
Fig.~\ref{fig:lunde}, which are crucial for the equilibration of the
system.  More precisely, the bosonic Hamiltonian
(\ref{eq:Luttinger_Hamiltonian}) provides an adequate description of
the excitation spectrum of a quantum liquid only at low energies,
namely $|\varepsilon|< D$, where the bandwidth $D\ll vp_F$.  Indeed,
for such excitations the spectrum can be linearized and consists of
two independent branches, as required in the Luttinger model.  On the
other hand, any excitation with energy $|\varepsilon|\sim vp_F$ is not
accounted for by the Hamiltonian (\ref{eq:Luttinger_Hamiltonian}).

This difficulty can be overcome as follows
\cite{matveev_equilibration_2012}.  Because the small probability of
an empty state near the bottom of the band plays the crucial role in
the physics of equilibration, let us first consider the spectrum of
the hole excitations.  For non-interacting fermions a hole with
momentum $Q$ can be defined as an excitation of the system obtained by
moving a fermion from state $p_F-Q$ to $p_F$.  For a system with
concave spectrum, such as the one in Fig.~\ref{fig:lunde}, the hole
represents the ground state of the system with the total momentum $Q$.
We use this observation to generalize the concept of a hole excitation
to the case of arbitrary interaction strength, and define the hole as
the ground state of the system with momentum $Q$.  Because moving a
fermion from one Fermi point to the other changes momentum by $2p_F$
without changing the energy of the system, the energy
$\varepsilon_Q^{}$ of the hole is a periodic function of momentum and
vanishes at $Q=0$, $\pm 2p_F$, $\pm4p_F$, \ldots.

\begin{figure}
\includegraphics[width=.35\textwidth]{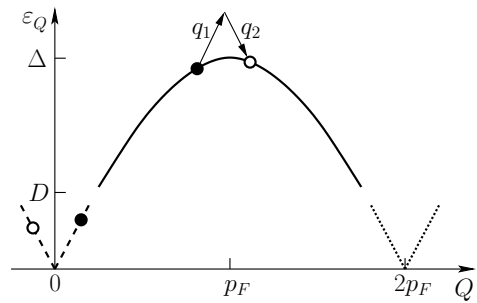}
\caption{Spectrum of a hole excitation in a quantum liquid.  The
  states with energies below $D$ are treated as excitations of the
  Luttinger liquid, whereas the higher-energy states are modeled as a
  mobile impurity.  The hole can change its momentum by $\delta
  Q=q_1-q_2$ by absorbing a boson with momentum $q_1$ and emitting one
  with momentum $q_2$.}
\label{fig:hole}
\end{figure}

The holes with energies below the bandwidth $D$ have a nearly linear
spectrum.  They are accounted for in the Hamiltonian
(\ref{eq:Luttinger_Hamiltonian}) as superpositions of various bosonic
excitations with the same momentum.  The holes with energies above $D$
are not included in the Hamiltonian (\ref{eq:Luttinger_Hamiltonian})
and are treated as mobile impurities in the Luttinger liquid
\cite{ogawa1992fermi, neto1996dynamics, pustilnik2006dynamic,
  khodas2007fermi,pereira2009spectral, imambekov_universal_2009,
  imambekov_phenomenology_2009, gangardt2010quantum,
  schecter_dynamics_2011}.  The exact value of the crossover energy
scale $D$ is not important as long as it is small compared to the
maximum energy of the hole $\varepsilon_{p_F}\sim vp_F$ and large compared
to the temperature $T$.

The mechanism of equilibration can be described as follows.  For
simplicity, we assume from now on that the liquid is at rest, $v_d=0$.
The gas of bosonic excitations equilibrates relatively quickly, and
the occupation numbers of bosonic states take the form
(\ref{eq:Boson_distribution}), which applies in the region $|q|<D/v$
represented by two straight dashed lines in Fig.~\ref{fig:hole}.  In a
generic case the total momentum of the excitations in the initial
state was not zero, so the Bose distribution
(\ref{eq:Boson_distribution}) has a boost velocity $u\neq0$.  As a
result of interactions between the bosons a small fraction of the
particles are promoted above the energy $D$, where they are no longer
described by the Hamiltonian (\ref{eq:Luttinger_Hamiltonian}).  At
arbitrary interaction strength the properties of these higher energy
excitations are rather complicated, but the lowest energy excitation
at a given momentum $Q$ is the hole.  Since $\varepsilon_Q\gg T$, the
occupation of the hole states is given by the Boltzmann factor
\begin{equation}
  \label{eq:Boltzmann_factor}
  f(Q)=e^{-(\varepsilon_Q-uQ)/T}.
\end{equation}
The presence of the correction $-uQ$ in the exponent is assured by the
fact that the hole interacts and exchanges momentum with the
thermalized bosons.  As a result of many such collisions the hole,
with small probability, may increase its momentum $Q$ above $p_F$,
after which it is more likely to fall toward $Q=2p_F$ than return to
the vicinity of $Q=0$, see Fig.~\ref{fig:hole}.  As the hole
approaches $Q=2p_F$, it enters the region of linear spectrum at
$\varepsilon_Q<D$, shown by dotted lines in Fig.~\ref{fig:hole}.
There it can again be viewed as a superposition of the bosonic
excitations.

As a result of this rare sequence of scattering events the bosons have
transferred to the hole the momentum $2p_F$.  Due to the conservation
of the total momentum (\ref{eq:Luttinger_Momentum}), this decrease of
momentum of the gas of excitations means that the zero mode
$J=N^R-N^L$ has increased by 2, i.e., one fermion has been
backscattered.  Also, the decrease of the total momentum of the bosons
means that the velocity $u$ has also decreased, in accordance with the
relaxation law (\ref{eq:relaxation}).

The equilibration proceeds very slowly because the
hole must pass the point $Q=p_F$ in momentum space, where the
occupation numbers are exponentially small.  One therefore expects
\begin{equation}
  \label{eq:rate_exponent}
  \tau^{-1} = C e^{-\Delta/T},
\quad
  \Delta=\varepsilon_{p_F}.
\end{equation}
To obtain the prefactor $C$, the kinetics of the scattering processes
should be considered in more detail.

We start by noting that the equilibration rate is controlled by a
small region of momentum space near $q=p_F$ where the energy of the
hole is close to the maximum, $\Delta-\varepsilon_Q\lesssim T$.  The
width of this region can be estimated as $(m^*T)^{1/2}$,
where we introduced the effective mass of the hole as
\begin{equation}
  \label{eq:mass}
  \frac{1}{m^*} = \left.-\frac{d^2\varepsilon_Q}{dQ^2}\right|_{Q=p_F}.
\end{equation}
Although the region is narrow compared to $p_F$, it is wide compared
to the typical change of the momentum of the hole in a single
collision with bosonic excitations.  Indeed, an elementary scattering
event consists of the hole absorbing one boson and emitting another,
see Fig.~\ref{fig:hole}.  Since the bosons are thermalized, the
typical change of $Q$ is of order $T/v$, which is much smaller than
$(m^*T)^{1/2}$ at $T\ll \Delta$.  This estimate enables us to simplify
the problem considerably.

The motion of the hole in momentum space is random and 
occurs in steps that are small compared to the size of the critical
region near the barrier.  Such diffusion in momentum space is
described by the Fokker-Planck equation \cite{lifshitz_physical_1981}
on the time-dependent distribution function $f(Q,t)$,
\begin{equation}
  \label{eq:Fokker-Planck}
  \partial_t f = -\partial_Q J,
\end{equation}
where the probability current $J$ has the form
\begin{equation}
  \label{eq:probability_current}
  J=-\frac{B(Q)}{2}\left[\frac{\varepsilon_Q'}{T}+\partial_Q\right]f.
\end{equation}
Here prime denotes derivative with respect to $Q$, and $B(Q)$ has the
meaning of the diffusion constant in momentum space.  It is defined as
\begin{equation}
  \label{eq:B_definition}
  B(Q)=\sum_{\delta Q}[\delta Q]^2 W_{Q,Q+\delta Q}
\end{equation}
in terms of the rate $W_{Q,Q'}$ of hole scattering from state
$Q$ to $Q'$.

In the steady state regime the Fokker-Planck equation
(\ref{eq:Fokker-Planck}) is solved by demanding that the probability
current $J$ be independent of $Q$.  To find the value of $J$ one has
to impose the boundary conditions on the occupation numbers $f(Q)$ on
the two sides of the barrier.  Considering that the size of the
crossover region in momentum space $(m^*T)^{1/2}$ is small compared
with $p_F$, one can approximate Eq.~(\ref{eq:Boltzmann_factor}) as
\begin{equation}
  \label{eq:boundary_condition_1}
  f(Q)=e^{-(\varepsilon_Q-up_F)/T},
\quad
  p_F-Q\gg (m^*T)^{1/2}.
\end{equation}
This expression specifies the boundary condition upon $f(Q)$ to the
left of the barrier.  To find one to the right of the barrier we
notice that the hole states with momenta $Q$ and $Q+2p_F$ are
identical, and the occupation of states with $Q$ between $p_F$ and
$2p_F$ is given by Eq.~(\ref{eq:Boltzmann_factor}) with $Q\to Q-2p_F$.
This yields
\begin{equation}
  \label{eq:boundary_condition_2}
  f(Q)=e^{-(\varepsilon_Q+up_F)/T},
\quad
  Q-p_F\gg (m^*T)^{1/2}.
\end{equation}
Solving the first-order differential equation
(\ref{eq:probability_current}) with constant $J$ one finds that the
boundary conditions (\ref{eq:boundary_condition_1}) and
(\ref{eq:boundary_condition_2}) are satisfied for
\begin{equation}
  \label{eq:J_result}
  J=u\,\frac{p_FB}{(2\pi m^*T^3)^{1/2}}\,e^{-\Delta/T},
\end{equation}
where we took the limit $u\to 0$ and denoted $B=B(2p_F)$.

A non-vanishing constant $J$ means that $(L/h)J$ holes are
passing any given point in momentum space in unit time.  Each hole
moving from the vicinity of $Q=0$ to that of $Q=2p_F$ takes momentum
$2p_F$ out of the bosonic excitations.  One therefore concludes that
the total momentum of the bosons changes with time at the rate $\dot
P_b=-2p_F (L/h)J$.  Given that the momentum of the bosons
distributed according to Eq.~(\ref{eq:Boson_distribution}) is
$P_b=(\pi L T^2/3\hbar v^3)u$, we find $\dot u=-u/\tau$ with
\begin{equation}
  \label{eq:rate_result}
  \frac1\tau=\frac{3B}{\pi^{5/2}p_F^2}
           \left(
             \frac{vp_F}{T}
           \right)^3
           \left(
             \frac{p_F^2}{2m^*T}
           \right)^{1/2}
           e^{-\Delta/T}.
\end{equation}
As expected, the equilibration rate has the exponential form
(\ref{eq:rate_exponent}).  To fully evaluate the prefactor, however,
one needs to study the hole scattering rate $W_{Q,Q'}$ and obtain the
diffusion constant (\ref{eq:B_definition}).

\section{Hole scattering rate}
\label{sec:hole-scattering-rate}

The scattering of a hole by bosonic excitations is a special case of
the problem of dynamics of a mobile impurity in a Luttinger liquid
\cite{neto1996dynamics}.  At low temperatures the leading scattering
process involves two bosons moving in the opposite directions.  By
absorbing one boson and emitting the other the impurity can scatter
from state $Q$ to a new state $Q'$ without violating conservation of
momentum and energy, Fig.~\ref{fig:hole}.  The authors of
Ref.~\onlinecite{neto1996dynamics} obtained the temperature dependence of
the mobility of the impurity in a Luttinger liquid in this regime,
$\mu\propto T^{-4}$.  Using the expression $\mu=T/B$ for the mobility  (see
\cite{lifshitz_physical_1981}, \S~21) one concludes
\begin{equation}
  \label{eq:B_vs_T}
  B=\chi T^5.
\end{equation}
The evaluation of the coefficient $\chi$ presents an interesting
problem.  Microscopic calculations can be performed in the special
cases of either weak or strong interactions
\cite{micklitz_transport_2010, matveev_equilibration_2010}.
Interestingly, one can also obtain a phenomenological expression for
$\chi$ in terms of the spectrum of the mobile impurity (hole) in the
Luttinger liquid \cite{matveev_equilibration_2012,
  matveev_scattering_2012}.  Here we review the latter approach.

The diffusion constant $B$ in the expression (\ref{eq:rate_result})
for the equilibration rate should be evaluated at $Q=p_F$.  On the
other hand, it is instructive to consider a more general problem and
study the hole scattering rate $W_{Q,Q+\delta Q}$ in
Eq.~(\ref{eq:B_definition}) for arbitrary $Q$.  The latter can be
found from the Fermi's Golden rule
\begin{eqnarray}
  W_{Q, Q+ \delta Q} &=& \frac{ 2 \pi }{ \hbar}
  \! \sum_{q_1,q_2} \!|t_{q_1,q_2}|^2 N_{q_1} (N_{q_2}+1)\,
  \delta_{q_1-q_2,\delta Q} 
  \nonumber \\
  && \times \delta (\varepsilon_Q - \varepsilon_{Q + \delta Q}
  +\hbar v |q_1| -\hbar v |q_2| ), 
\label{eq:golden_rule}
\end{eqnarray}
where $t_{q_1,q_2}$ is the matrix element of the process in which the
hole absorbs the boson $q_1$ and emits the boson $q_2$,
Fig.~\ref{fig:hole}.  Since the typical energies of the bosons are of
order temperature, we will assume $|\delta Q|\ll p_F$.  In this case
one can easily obtain the momenta $q_1$ and $q_2$ from the
conservation laws,
\begin{equation}\label{eq:q_Q}
    q_1=\frac12\delta Q +\frac{v_Q}{2 v}|\delta Q|, 
\quad 
    q_2=-\frac12\delta Q +\frac{v_Q}{2 v}|\delta Q|.
\end{equation}
Here $v_Q^{}=\varepsilon_Q'$ is the velocity of the hole with momentum
$Q$.  Using Eq.~(\ref{eq:q_Q}) one easily expresses the scattering
rate as
\begin{equation}
  \label{eq:W_from_golden_rule}
  W_{Q, Q+ \delta Q} = \frac{L}{\hbar^2 v}\,N_{q_1} (N_{q_2}+1)|t_{q_1,q_2}|^2.
\end{equation}
To evaluate the matrix element $t_{q_1,q_2}$ we need to discuss the
Hamiltonian of the Luttinger liquid in the presence of a mobile
impurity.

We start by writing the Hamiltonian (\ref{eq:Luttinger_Hamiltonian})
in an alternative form \cite{giamarchi2004quantum}
\begin{equation}\label{eq:H_0}
    H_0=\frac{\hbar v}{2\pi} 
        \int dx[ K (\nabla \theta)^2 +K^{-1}(\nabla \phi)^2],
\end{equation}
where the two bosonic fields $\phi(x)$ and $\theta(x)$ satisfy the
commutation relation
\begin{equation}\label{eq:commutation_phi_theta}
    [\phi(x), \nabla\theta(x')]=i \pi \delta(x-x')
\end{equation}
and the Luttinger liquid parameter $K$ depends on the interactions
between particles.  The case of non-interacting fermions corresponds
to $K=1$.

The Hamiltonian (\ref{eq:H_0}) can be brought to the form
(\ref{eq:Luttinger_Hamiltonian}) with the help of the following
expressions for the fields $\phi$ and $\theta$ in terms of the bosonic
operators: 
\begin{eqnarray}
  \label{eq:bosons_phi}
 \!\!\!\!\!\!\! 
\nabla \phi (x) \!&\!=\!&-i\sum_q\sqrt{\frac{\pi K|q|}{2\hbar L}}\, 
       \mathrm{sgn} (q)
       (b_q+b_{-q}^\dagger)e^{iqx/\hbar},
\\
  \!\!\!\!\!\!\!\nabla \theta(x)\!&\!=\!& i \sum_q\sqrt{\frac{\pi |q|
    }{2\hbar K L}}\,
       (b_q-b_{-q}^\dagger)e^{iqx/\hbar}.
\label{eq:bosons_theta}
\end{eqnarray}
The advantage of the form (\ref{eq:H_0}) of the Hamiltonian is that
the fields $\phi$ and $\theta$ have clear meanings in terms of the
observables characterizing the quantum liquid.  For instance, the
field $\phi(x)$ accounts for the fluctuations of the density of the
liquid
\begin{equation}\label{eq:phi_density}
    n(x)=n_0 +\frac{1}{\pi} \nabla \phi(x),
\end{equation}
where $n_0=N/L$ is the average density \cite{giamarchi2004quantum}.
Similarly, the field $\theta$ is related to the momentum $\kappa$ of
the liquid per particle
\begin{equation}
  \label{eq:theta_kappa}
   \kappa (x)= - \hbar  \nabla \theta (x),
\end{equation}
see, e.g., Ref.~\onlinecite{matveev_scattering_2012}.

The coupling of the hole to bosonic excitations in the Luttinger
liquid can now be obtained by considering the dependence
$\varepsilon_Q(n,\kappa)$ of the energy of the hole on the density and
momentum of the liquid.  Using Eqs.~(\ref{eq:phi_density}) and
(\ref{eq:theta_kappa}) we expand $\varepsilon_Q(n,\kappa)$ in powers
of the bosonic fields,
\begin{eqnarray}
  \label{eq:epsilon_Q_expansion}
  \varepsilon_Q(n,\kappa) &=& 
     \varepsilon_Q(n_0,0)
     +\frac{1}{\pi}\partial_n\varepsilon_Q\nabla\phi
     -\hbar\,\partial_\kappa\varepsilon_Q\nabla\theta
\nonumber\\
   &&+\frac{1}{2\pi^2}\partial_n^2\varepsilon_Q(\nabla\phi)^2
     +\frac{\hbar^2}{2}\partial_\kappa^2\varepsilon_Q(\nabla\theta)^2
\nonumber\\
   &&-\frac{\hbar}{\pi}
      \partial_n\partial_\kappa\varepsilon_Q\nabla\phi\nabla\theta+\ldots
\end{eqnarray}
Here all derivatives of $\varepsilon_Q(n,\kappa)$ are taken at $n=n_0$
and $\kappa=0$.  Taking into account Eqs.~(\ref{eq:bosons_phi}) and
(\ref{eq:bosons_theta}) one sees that the second-order terms in
Eq.~(\ref{eq:epsilon_Q_expansion}) contain contributions in which a
boson $q_1$ is absorbed and boson $q_2$ on the opposite branch is
emitted.  The respective matrix element has the form
\begin{equation}
  \label{eq:t-a}
  t_{q_1,q_2}^{(a)}=-\frac{\sqrt{|q_1 q_2|}}{2\pi\hbar L}\,
                  \partial^2_{LR}\varepsilon_Q,
\end{equation}
where we assumed that the hole is at $x=0$ and introduced the notation
\begin{equation}
  \label{eq:partial_LR}
  \partial^2_{LR} = K\partial_n^2-\frac{(\pi\hbar)^2}{K}\partial_\kappa^2.
\end{equation}

In addition to the terms coupling the hole to two bosons,
Eq.~(\ref{eq:epsilon_Q_expansion}) contains the contribution linear in
bosonic operators:
\begin{equation}
  \label{eq:linear_coupling}
  i\partial_L\varepsilon_Q\sum_{q<0} \sqrt{\frac{|q|}{2\pi\hbar L}}\,
     (b_q-b_q^\dagger)
  -i\partial_R\varepsilon_Q\sum_{q>0} \sqrt{\frac{|q|}{2\pi\hbar L}}\,
     (b_q-b_q^\dagger),
\end{equation}
where
\begin{equation}
  \label{eq:partials}
  \partial_L=\sqrt{K}\partial_n
             -\frac{\pi\hbar}{\sqrt K}\partial_\kappa,
\quad
  \partial_R=\sqrt{K}\partial_n
             +\frac{\pi\hbar}{\sqrt K}\partial_\kappa.
\end{equation}
The linear coupling terms (\ref{eq:linear_coupling}) also contribute
to the matrix element $t_{q_1,q_2}$, but in the second order
perturbation theory,
\begin{eqnarray}
  \label{eq:t-b_initial}
  t_{q_1,q_2}^{(b)}&=&-\frac{\sqrt{|q_1 q_2|}}{2\pi\hbar L}
     \left[
      \frac{\partial_L\varepsilon_{Q+q_1}\partial_R\varepsilon_Q}
           {\varepsilon_Q+vq_1-\varepsilon_{Q+q_1}}
     \right.
\nonumber\\
     &&\hspace{5em}\left.
        +\frac{\partial_R\varepsilon_{Q-q_2}\partial_L\varepsilon_Q}
            {\varepsilon_Q-\varepsilon_{Q-q_2}-v|q_2|}
     \right],
\end{eqnarray}
where we assumed $q_1>0$ and $q_2<0$, i.e., positive $\delta Q$ as in
Fig.~\ref{fig:hole}.  It is important to account for the corrections
to the momentum of the hole in the numerator which are due to the fact
that the two perturbations of the form (\ref{eq:linear_coupling}) act
on the states of the hole with different values of $Q$.

The expression (\ref{eq:t-b_initial}) can be simplified by taking
advantage of the smallness of $q_1\sim q_2\sim\delta Q\ll p_F$.  The
expression in the brackets appears to scale as $1/\delta Q$.  This
term is obtained by neglecting corrections to $Q$ in the numerator and
linearizing the denominators in $q_1$ and $q_2$.  However, for the
specific values (\ref{eq:q_Q}) of the boson momenta the two
contributions in the brackets cancel each other.  Evaluating the next
order terms in $q_1$ and $q_2$ one finds
\begin{eqnarray}
  \label{eq:t-b}
\!\!\!\!\!\!\!\!\!
  t_{q_1,q_2}^{(b)}&\!=\!&-\frac{\sqrt{|q_1 q_2|}}{2\pi\hbar L}
      \Bigg[
      \frac{1}{m_Q^*}
      \frac{\partial_L\varepsilon_{Q}}{v+v_Q}
      \frac{\partial_R\varepsilon_{Q}}{v-v_Q}
\nonumber\\
   &&\hspace{4em}
     +\partial_L v_Q
      \frac{\partial_R\varepsilon_{Q}}{v-v_Q}
     -\partial_R v_Q
      \frac{\partial_L\varepsilon_{Q}}{v+v_Q}\Bigg].
\end{eqnarray}
Here the momentum dependent effective mass of the hole is defined by
$1/m_Q^*=-\varepsilon_Q''$.  The first term in Eq.~(\ref{eq:t-b})
originates from the expansion of the denominators of
Eq.~(\ref{eq:t-b_initial}) to second order in $q_1$ and $q_2$, whereas
the remaining two terms are obtained by accounting for linear
corrections in the numerators.

Finally, one more contribution to the scattering matrix element
$t_{q_1,q_2}$ is obtained when the hole couples to a single boson,
Eq.~(\ref{eq:linear_coupling}), which in turn splits into two.  The
latter matrix element involves three bosons and should therefore
originate from cubic in $\phi$ and $\theta$ corrections in the
Hamiltonian.  For a fluid at rest the symmetry allows only for even
powers of $\theta$, so the correction must have the form
\begin{equation}
    H_\alpha=\int d x \big[\alpha_\theta ( \nabla \phi )(\nabla\theta)^2 
        + \alpha_\phi (\nabla \phi)^3\big].
\label{eq:H_alpha_Euler}
\end{equation}
The values of the coefficients $\alpha_\theta$ and $\alpha_\phi$ can
be related to the density dependences of the parameters $v$ and $K$ of
the quadratic Hamiltonian (\ref{eq:H_0}) by considering the correction
to the total Hamiltonian $H_0+H_\alpha$ caused by a small change of
particle density $\delta n$.  As a result one obtains
\cite{matveev_scattering_2012}
\begin{equation}
\label{eq:alpha_relations}
  \alpha_\theta = \frac{\hbar}{ 2\pi^2}\,  \partial_n(v K), 
\quad
  \alpha_\phi = \frac{\hbar}{6\pi^2}\, \partial_n \left( \frac{v}{K}\right).
\end{equation}
In order to find a contribution to $t_{q_1,q_2}$ we need the matrix
element of $H_\alpha$ that absorbs a boson with momentum $q_1$ on one
branch and emits a boson $q_2$ on the other branch.  Using
Eq.~(\ref{eq:alpha_relations}) we obtain
\begin{equation}
  \label{eq:three_bosons}
   \frac{i\mathrm{sgn} (q_1)}{\sqrt{2\pi\hbar L}}
   \frac{v\partial_n K}{\sqrt{K}}
   \sqrt{|q_1 q_2(q_1-q_2)|}\,
   b_{q_2}^\dagger b_{q_1} \big(b_{q_2-q_1}+b_{q_1-q_2}^\dagger\big).
\end{equation}
The second order calculation of the matrix element $t_{q_1,q_2}$ with
perturbations (\ref{eq:three_bosons}) and (\ref{eq:linear_coupling})
yields
\begin{equation}
  \label{eq:t-c}
  t_{q_1,q_2}^{(c)} = -\frac{\sqrt{|q_1 q_2|}}{2\pi\hbar L}
                   \frac{v\partial_n K}{\sqrt K}
                   \bigg(
                    \frac{\partial_R\varepsilon_Q}{v-v_Q}
                    +\frac{\partial_L\varepsilon_Q}{v+v_Q}
                   \bigg).
\end{equation}

\begin{figure}
\includegraphics[width=.45\textwidth]{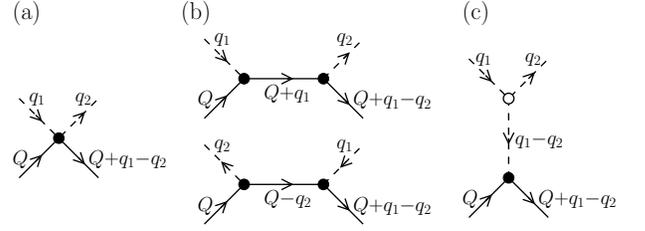}
\caption{The three types of processes contributing to the scattering
  matrix element $t_{q_1,q_2}$. (a) The first order process, in which
  the hole couples to two bosons.  (b) The second order processes
  involving two perturbations, each coupling the hole to one of
  the two bosons.  (c) The second order process where the hole
  couples to a virtual boson, which is separately coupled to bosons $q_1$
  and $q_2$.}
\label{fig:three_processes}
\end{figure}

The three types of processes leading to hole scattering with
absorption of boson $q_1$ and emission of boson $q_2$ are
illustrated in Fig.~\ref{fig:three_processes}.  Their total is given by
\begin{equation}
  \label{eq:t-total}
    t_{q_1,q_2} = -\frac{\sqrt{|q_1 q_2|}}{2\pi\hbar L}\,Y_{Q},
\end{equation}
where
\begin{eqnarray}
  \label{eq:Y}
      &&Y_Q=
      \partial^2_{LR}\varepsilon_Q+
      \frac{1}{m_Q^*}
      \frac{\partial_L\varepsilon_{Q}}{v+v_Q}
      \frac{\partial_R\varepsilon_{Q}}{v-v_Q}
     +\partial_L v_Q
      \frac{\partial_R\varepsilon_{Q}}{v-v_Q}
\nonumber\\
   &&\hspace{0.5em} -\partial_R v_Q
      \frac{\partial_L\varepsilon_{Q}}{v+v_Q}
      +\frac{v\partial_n K}{\sqrt K}
      \bigg(
       \frac{\partial_R\varepsilon_Q}{v-v_Q}
       +\frac{\partial_L\varepsilon_Q}{v+v_Q}
      \bigg).
\end{eqnarray}
An alternative way of evaluating the scattering matrix element
involves performing a unitary transformation that eliminates the
linear coupling (\ref{eq:linear_coupling}) of the hole to the bosons
\cite{matveev_scattering_2012}.  Upon this transformation only the
quadratic coupling remains, which is then evaluated in the first
order, similarly to Eq.~(\ref{eq:t-a}).  The resulting expression given
by Eqs. (49) and (50) of Ref.~\onlinecite{matveev_scattering_2012} is
equivalent to Eq.~(\ref{eq:t-total}).

Using the expression (\ref{eq:t-total}) for the scattering matrix
element in combination with Eqs.~(\ref{eq:W_from_golden_rule}) and
(\ref{eq:B_definition}) one easily recovers the temperature dependence
(\ref{eq:B_vs_T}).  The coefficient $\chi$ takes the form
\begin{equation}
  \label{eq:chi}
  \chi=\frac{4\pi Y_{p_F}^2}{15\hbar^5 v^6}.
\end{equation}
Equations (\ref{eq:rate_result}), (\ref{eq:B_vs_T}), and
(\ref{eq:chi}) provide a complete expression for the equilibration
rate of a one-dimensional quantum liquid in terms of the spectrum of
hole excitations and its dependences on the particle density $n$ and
the momentum per particle $\kappa$.

\section{Discussion}
\label{sec:discussion}

In this paper we discussed the equilibration of a one-dimensional
quantum liquid of interacting fermions.  The conventional Luttinger
liquid theory \cite{haldane1981luttinger, giamarchi2004quantum} of
these systems neglects the processes of backscattering.  In many cases
this is an excellent approximation since the corresponding scattering
rates are exponentially small at low temperatures,
Eq.~(\ref{eq:rate_result}).  However, the Luttinger liquid
approximation does not enable one to treat a number of interesting
phenomena in which the backscattering plays the crucial role.

One example is the conductance of a long uniform quantum wire.  The
Luttinger liquid theory predicts perfect conductance quantization in
these devices, regardless of the interaction strength
\cite{maslov_landauer_1995, ponomarenko_renormalization_1995,
  safi_transport_1995}.  On the other hand, it is easy to show that at
weak electron-electron interactions a correction to conductance
appears due to the backscattering processes
\cite{lunde_three-particle_2007, micklitz_transport_2010}.
Interestingly, an expression for the conductance of a quantum wire can
be obtained for any interaction strength
\cite{matveev_equilibration_2011},
\begin{equation}
  \label{eq:conductance}
  G=\frac{e^2}{h}\left(1-\frac{\pi^2}{3}\,
     \frac{T^2}{v^2p_F^2}\frac{L}{L+2v\tau}\right).
\end{equation}
The backscattering gives rise to a negative correction to the
quantized conductance, which grows with temperature and with the
length of the wire $L$.  In short wires the correction $\delta
G\propto \tau^{-1}$ is exponentially small, but it saturates at
$\delta G\sim -(e^2/h)(T/vp_F)^2$ in long wires.

Temperature dependent corrections to conductance of quantum wire
devices have been observed in multiple experiments
\cite{thomas_possible_1996, kristensen_bias_2000,
  cronenwett_low-temperature_2002}.  The data shows excellent
quantization of conductance at lowest temperatures and a negative
correction developing as the temperature is raised.  These
observations are in qualitative agreement with
Eq.~(\ref{eq:conductance}).  In comparing the data with theory it is
important to keep in mind that our discussion so far has ignored
spins, which appear to play an important role in experiments.  The
result (\ref{eq:conductance}) can be generalized to include spins
\cite{matveev_equilibration_2011}, but the evaluation of the
equilibration rate of a system with spins is still an open problem.
Another complication is that most experiments study rather short wires
which may not be treated as uniform.

Our discussion of the equilibration rate did not assume Galilean
invariance of the system.  On the other hand, momentum conservation
was assumed.  Thus the results do not automatically apply to systems
of interacting particles in periodic potentials, such as spin chains.
In such systems umklapp scattering by the external potential may
facilitate equilibration.  On the other hand, electrons in GaAs
quantum wires have an essentially quadratic spectrum
$\epsilon_p=p^2/2m$, where $m$ is the effective mass of electron in
this material.  Such electron system is Galilean invariant, which
leads to a few simplifications.  First, the Luttinger liquid parameter
in this case is determined by velocity of the bosons, $K=\pi\hbar
n/mv$.  Second, the dependence of the excitation energy on momentum
$\kappa$ has the simple form \cite{lifshitz_statistical_1980}
\begin{equation}
  \label{eq:Galilean_transformation}
    \varepsilon_Q(n, \kappa) =\varepsilon_Q(n) +  Q \frac{\kappa }{m}.
\end{equation}
For momenta $Q$ in the vicinity of $p_F$ one can expand
$\varepsilon_Q(n)=\Delta(n) -(Q-p_F)^2/2m^*$ and find
\begin{equation}
  \label{eq:Y_Galilean}
  Y_{p_F}=K\left(
           \Delta''+ \frac{\Delta'^2}{m^* v^2} -\frac{2v'}{v}\Delta'
          \right).
\end{equation}
Upon substitution into Eq.~(\ref{eq:chi}) one recovers the results of
Ref.~\onlinecite{matveev_equilibration_2012} for equilibration rate in
Galilean invariant systems obtained by a different technique.  It is
worth noting that in this case the equilibration rate is fully
determined by the density dependences of the velocity of bosonic
excitations $v$ and the maximum energy of the hole $\Delta$.

Although our main focus was on interacting systems of fermions, the
approach and the results should be equally applicable to systems of
bosons.  Similar techniques have been recently applied to dynamics of
dark solitons and mobile impurities in bosonic fluids
\cite{gangardt2010quantum,schecter_dynamics_2011}.  Finally, it is
worth mentioning that in integrable models apart from energy and
momentum there are multiple additional conserved quantities, and one
expects that no equilibration of the system should take place.  In
particular one should find $\tau^{-1}=0$.  This conjecture has been
checked \cite{matveev_equilibration_2012} for the Calogero-Sutherland
\cite{sutherland2004beautiful} and Lieb-Liniger \cite{lieb_exact_1963}
models.  More generally, one expects \cite{matveev_scattering_2012}
that for integrable models the quantity $Y_Q$ given by
Eq.~(\ref{eq:Y}) should vanish for any $Q$.

\begin{acknowledgments}
  The author is grateful to A.~V. Andreev and A. Furusaki for
  discussions and to RIKEN for kind hospitality.  Work supported by
  UChicago Argonne, LLC, under contract {No.} DE-AC02-06CH11357.
\end{acknowledgments}


\end{document}